\RequirePackage{fix-cm}
\documentclass{svjour3}                     
\usepackage{amsmath}
\usepackage{amssymb}
\usepackage{amsfonts}
\usepackage{graphicx}
\usepackage{epstopdf}
\usepackage{color}
\usepackage{soul}
\usepackage[caption=false]{subfig}
\usepackage[toc,page]{appendix}
\usepackage{multirow}
\usepackage[section]{placeins}
\usepackage{physics}
\usepackage{comment}

\graphicspath{{paper_plots/}}

\begin{document}
%
\title{Multistable Synaptic Plasticity induces Memory Effects and Cohabitation of Chimera and Bump States in Leaky Integrate-and-Fire Networks}

\author{A. Provata \and Y. Almirantis \and W. Li}
\institute{
A. Provata \at Institute of Nanoscience and Nanotechnology, National Center for Scientific Research ``Demokritos'', 15341 Athens, Greece, \email{a.provata@inn.demokritos.gr }. \\
Y. Almirantis \at Institute of Nanoscience and Nanotechnology, National Center for Scientific Research ``Demokritos'', 15341 Athens, Greece,
\email{yalmir@bio.demokritos.gr  }. \\
W. Li \at Department of Applied Mathematics and Statistics, Stony Brook University, Stony Brook, NY, USA, and
The Robert S. Boas Center for Genomics and Human Genetics,
The Feinstein Institutes for Medical Research, Northwell Health, Manhasset, NY, USA,
\email{wtli2012@gmail.com}.
}

\date{Received: \today / Revised version: date}
\maketitle
%
\begin{abstract}
{
Chimera states and bump states are collective synchronization phenomena observed independently (at different parameter regions) in networks of coupled nonlinear oscillators. And while chimera states are characterized by coexistence of coherent and incoherent domains, bump states consist of active domains operating on a silent background. Multistable plasticity in the network connections originates from brain dynamics and is based on the idea that neural cells may transmit inhibitory or excitatory signals depending on various factors, such as local connectivity, influence of neighboring cells etc. During the system/network integration, the link weights adapt and, in the case of multistability, they may organize in coexisting excitatory and/or inhibitory domains. Here, we explore the influence of bistable plasticity on collective synchronization states and we numerically demonstrate that the dynamics of the linking may give rise to co-existence of bump-like and chimera-like states simultaneously in the network. In the case of bump and chimera co-existence, confinement effects are developed: the different domains stay localized and do not travel around the network. Memory effects are also reported  in the sense that the final spatial arrangement of the coupling strengths reflects some of the local properties of the initial link distribution. For the quantification of the system's spatial and temporal features, the global and local entropy functions are employed as measures of the network organization, while the average firing rates account for the network evolution and dynamics.
}
\end{abstract}

\keywords{synchronization; multistable plasticity; synaptic plasticity; weighted network; 
adaptive network; chimera states; bump states; memory effects; confinement; leaky Integrate-and-Fire model.}
\maketitle
\section{Introduction}
\label{sec:intro}
\par During the past two decades considerable efforts have been devoted to the understanding
of complex synchronization phenomena observed in networks of coupled nonlinear
oscillators. Such phenomena include traveling waves, spiral (2D) and scroll (3D) waves, chimera states and bump states
to name just a few\cite{panaggio:2015,schoell:2016,oomelchenko:2018,majhi:2019,zakharova:2020}.
In particular, the phenomena of hybrid synchronization (chimera states or bump states)
were unexpected since 
they were first shown to occur in networks consisting of identical elements with
identical linking  between nodes and were, therefore, associated 
with spatial symmetry breaking in the network \cite{kuramoto:2002,abrams:2004}. While in the earlier works the 
connectivity in the system was considered to be constant/uniform or random or statistically
homogeneous, in recent years the research interest is shifted toward temporal and adaptive networks 
whose link weights depend on time and they may gradually develop
heterogeneous connectivity properties. Drawing inspiration by the recent advances in the domain
of temporal neural networks we here
investigate the influence of link adaptivity in the emergence and the form of hybrid
synchronization patterns. 

\par From the point of view of applications, 
the studies of coupled nonlinear oscillators are motivated by brain dynamics,
where  neural cells operate as potential integrators and they connect with other neural cells
via their axons forming large complex networks \cite{gerstner:2002,kandel:2013,haddad:2014,brodal:2016}.
The intricate structure of these networks allows the propagation of information in the brain
for the control and regulation of a wide variety of body and cognitive functions 
such as sensory processes, motor actions, thinking, problem-solving, 
reasoning, learning, memory, emotions, consciousness, homeostasis and many others. 
\par Recent studies of the network structure of the human brain
have shown that the connectivity between neural cells is not static but dynamical and
changes with time due to aging or diseases or due to the need for
adaption to a constantly changing potential environment \cite{wilson:1973,ewald:2012,neuner:2014,andreou:2017,butz:2009}. 
This adaptation process is mainly achieved via plasticity of the connectivity weights.
The adaptation of the weights to the local potential environment or to external stimuli
is regarded as a learning processes, meaning that the parameters (connectivity weights)
in the network are adjusted to optimize the organism's 
ability to survive, to copy with new challenges, to develop and to respond to life's demands in general.
\footnote{As will be noted later, in the present work adaptivity and plasticity will be considered with respect to 
the local weight environment and not to the local potential environment used mainly in the literature.}
Besides their relevance in life sciences, temporal and adaptive neural networks are now intensively studied in
connection with the recent advances in machine learning and artificial intelligence \cite{breakspear:2017}. 
Motivated by the plasticity of the connectivity weights in brain dynamics, we study here the influence
of adaptivity in the formation of hybrid synchronization states, such as chimera states
and bump states, using as exemplary dynamics the leaky Integrate-and-Fire (LIF) model.

\par When studying adaptive connectivity, we keep in mind that there will be an heterogeneity
in the coupling weights of the network after adaptation, even if we start from homogeneous
or uniformly randomly distributed initial weights. As the original definition of chimera  states is
based on identical dynamics and identical linking of all nodes \cite{omelchenko:2013,omelchenko:2015}, 
the patterns resulting due to adaptivity, if composed by co-existing of coherent and incoherent domains
in networks with heterogeneous weights, will be called termed ``chimera-like'' states.
Likewise,``bump-like'' states will be states composed by coexisting active and subthreshold
(silent) domains observed in adaptive
networks where the linking between the nodes may not be homogeneous.

\par Previews studies of complex synchronization phenomena in LIF networks have revealed the presence
of chimera states in a 1D ring, 2D toroidal and 3D hypertoroidal geometries under nonlocal connectivity
and for relatively large values of the coupling ranges.
Studies on 1D ring geometries demonstrate that chimera and multichimera states are possible 
depending on parameters such as coupling range and coupling strength and on the connectivity schemes
(including hierarchical, fractal connectivities)
\cite{olmi:2010,olmi:2015,tsigkri:2016,tsigkri:2018,politi:2018,olmi:2019b}. 
In 2D toroidal geometries, striking chimera patterns were revealed in the form of coherent or
incoherent spots, stripes, grids of spots, spiral waves and other composite patterns \cite{schmidt:2017,argyropoulos:2019a}.
In 3D hypertorus geometries the chimera forms were generalized to coherent and incoherent spheres,
cylinders, planes and  composite patterns \cite{kasimatis:2018}.
Concerning the presence of bump states in LIF networks, 
traveling and spatially confined bumps have been reported in 1D and 2D geometries.
The multiplicity of bumps, their size and
traveling speed have been shown to 
depend on the parameter values (coupling strength, coupling range, refractory period, density of inactive nodes etc.)
\cite{laing:2001,laing:2020,laing:2021,oomelchenko:2021,argyropoulos:2019a,tsigkri:2017,provata:2024b}.

\par The chimera and bump states recapitulated in the previous paragraph were all obtained using
identical LIF elements on all nodes of the network and identical non-local linking with constant and
equal weights. To keep the connectivity geometry identical in all nodes and to avoid boundary value
effects all simulations were performed with periodic boundary conditions in one, two or three dimensions.
In the present study, we investigate the influence of link adaptivity on the formation of chimera/bump
states using as an exemplary case bistable couplings evolving via diffusive non-local interactions 
\cite{oomelchenko:2018,omelchenko:2013}.

\par In earlier studies multistability was introduced already in the structural
parameters of the single node dynamics. Such attempts include bistability
in the internal frequency \cite{provata:2020} or in the amplitude \cite{provata:2024} of the oscillators.
As a result complex frequency and amplitude chimera states were produced
as well as frequency-amplitude entanglement effects  \cite{provata:2024}.
In the present study we introduce multistability in the bonding between
the different nodes in the network. Synaptic multistability
may be induced by the neuron-glial interaction in the brain \cite{lazarevich:2017}
and its biological meaning is that
 the strength of the connections  between neurons (via the axons)
may not be an altogether continuous function but may take specific values, 
excitatory or inhibitory, leading initially random coupling values to multiple
fixed points. As a result, different domains in the system are governed
by different coupling parameters and these influence the local firing rates
and the internal dynamics. We will demonstrate in the sequel that the
effects of introducing bistability in the coupling strength may lead,
for specific parameter domains, into coexistence of bumps and chimeras
in the same network. Other interesting effects include a) the shift of the 
coupling parameters where non-trivial composite chimera-like or bump-like dynamics are observed,
b) memory effects in the sense that the final spatial distribution of coupling strengths recalls traces of their initial
distribution and
c) the presence of a transition point in the coupling
range where bistability terminates abruptly giving rise to single fixed point oscillatory dynamics. 

\par We should stress here that the bistable adaptivity rules proposed here
are not to be confused with Hebbian like adaptivity, where plasticity depends on the 
pre- and post-synaptic potentials following the Hebb's principle 
that ``neurons that fire together, wire together'' 
or other spike-timing-dependent plasticity (STDP) models \cite{hebb:1949,choi:2022,nunzio:2022,berner:2023}.
Rather, in the present study adaptivity/plasticity is inherent in the structure
of the neuron axons which are considered to be bistable (or more generally, multistable) and they adapt 
according to the general tendencies of link weights in the neighboring 
neural environment \cite{lazarevich:2017,holme:2012}.

\par
In the next section we present the single/uncoupled LIF model, the ring LIF network and the plasticity rules employed.
In Sec.~\ref{sec:chimera_plasticity} we study the formation of complex chimera-like states under bistability in the link
plasticity and 
in Sec.~\ref{sec:bumps_plasticity} we examine the influence of link plasticity on the formation
of bump-like states. In both  Secs.~\ref{sec:chimera_plasticity}
and~\ref{sec:bumps_plasticity}  we provide evidence and  stress the presence of spatial memory effects.
In Sec.~\ref{sec:multistable_plasticity} we investigate the effects that link plasticity induces
on the network dynamics; namely, we show that multistability on the link weights causes cohabitation
of chimera states and bump states on the ring network. 
In Sec.~\ref{sec:coupling-range} entropy values are used for the quantification of the network dynamics.
Using the local entropy values as quantitative index, we show that for large coupling
ranges the system transits from bistable to homogeneous, while the transition $R$-values depend
on the coupling parameters.
 In the Conclusions section we recapitulate our main results and discuss
open problems. To facilitate the reading of this manuscript and to highlight the influence of
plasticity in the network, in the two Appendix sections we recapitulate some results on
the presence of chimera or bump states in LIF networks without plasticity rules.

\section{The model}
\label{sec:model}

\par In this section we present the LIF coupling scheme and its implementation in a 1D ring geometry
 as will be used in the simulations.
 The difference of this approach from previous studies in the literature lies in the choice
of the coupling terms, which here become time-dependent and adaptive to the neighboring coupling environment.

\subsection{The uncoupled LIF  dynamics}
\par Historically, the first Integrate-and-Fire neuron models 
were introduced by Louis Lapicque at the turn of the 20th century
to describe the response of nerve fibers to electrical stimuli \cite{lapicque:1907,lapicque:1907a,lapicque:1907b}.
The variable $u(t)$ introduced by Lapicque describes the neuron membrane potential which grows linearly due to the interaction
with other neurons, up to a threshold potential, $u_{\rm th}$. After reaching $u_{\rm th}$
 the $u$-variable (membrane potential) is automatically reset
to its rest state value, $u_{\rm rest}$, forming a periodically firing oscillator. 
A leaky term (proportional to the potential $u_i(t)$ ) was introduced to inherently avoid potential divergences
in the long time scales  and  to match experimental observations.
The version which includes the leaky term is called ``Leaky Integrate-and-Fire'' neuron model or simply LIF. 
The various variations of integrate-and-fire models are very
popular amongst computational neuroscientists due to their simple dynamics allowing to consider many thousands
up to millions of coupled neurons to mimic realistic natural systems. The choice of the variant depends on the
particular application. In the present study the LIF model is employed and the dynamics
of the single LIF neural oscillator reads as: 

\begin{subequations}
\begin{equation}
\label{eq01a} 
\frac{du(t)}{dt}= \mu - u(t)
\end{equation}
\begin{equation}
\lim_{\epsilon \to 0}u(t+\epsilon ) \to u_{\rm rest}, 
\>\>\> {\rm when} \>\> u(t) \ge u_{\rm th}.
\label{eq01b}
\end{equation}
\label{eq01}
\end{subequations}
\par Within the interval $\left[ u_{\rm rest}, u_{\rm th} \right]$ the solution
to Eq.~\eqref{eq01} can be obtained analytically and the period $T_s$ that the
single LIF oscillator takes to fire after rest is calculated as
\begin{eqnarray}
T_s =\ln \left[ \frac{ \mu - u_{\rm rest}}{  \mu - u_{\rm_{th}}}\right] .
\label{eq02}
\end{eqnarray}

\subsection{The coupled and adaptive LIF  dynamics}
\label{sec:adaptive}
\par When many LIF elements are coupled on a ring network,
the differential equations which describe the evolution of the potential $u_{i}(t)$ 
of a neuron at
position $i$ on the ring of $N$ neurons are provided below in Eq.~\eqref{eq03a} and~\eqref{eq03b}. 
Because the coupling strengths are also time dependent, an additional equation, Eq.~\eqref{eq03c},
is necessary to describe the evolution of the coupling strengths, $\sigma_i(t),\>\> i=1,N$.

\begin{subequations}
\begin{equation}
\label{eq03a} 
\frac{du_{i}(t)}{dt}= \mu - u_{i}(t)+\frac{1}{2R}\sum_{j=i-R }^{i+R} \sigma_{i}(t)
\left[ u_{j}(t) - u_{i}(t)\right]
\end{equation}
\begin{equation}
\lim_{\epsilon \to 0}u_{i}(t+\epsilon ) \to u_{\rm rest}, 
\>\>\> {\rm when} \>\> u_{i}(t) \ge u_{\rm th}.
\label{eq03b}
\end{equation}
\begin{equation}
\label{eq03c} 
\frac{d\sigma_{i}(t)}{dt}= c_{\sigma}(\sigma_i - \sigma_l) (\sigma_i - \sigma_c) (\sigma_i - \sigma_h)
+\frac{s}{2R}\sum_{j=i-R }^{i+R} \left[ \sigma_{j}(t) - \sigma_{i}(t)\right]
\end{equation}
\label{eq03}
\end{subequations}

\par In Eqs.~(\ref{eq03}) the parameter $u_{\rm th}$ is identical for all neurons:
when reaching it they reset to their rest potential $u_{\rm rest}$. 
 $ \mu $ is the value that the potential of any neuron would asymptotically tend to, if there was 
no resetting condition; therefore, to achieve periodic firing $u_{\rm th}$ must satisfy the condition $u_{\rm th} < \mu $.
The interaction kernel is assumed to be of rectangular form and 
every neural oscillator $i$ is linked with all other oscillators $j$ in the range
$i-R \le  j \le i+R$  with adaptive coupling 
strength $\sigma_{i}(t)$. 
\par The coupling strengths in Eq.~\eqref{eq03c} are assumed to evolve in time following a 
nonlinear law of order three.
The value of the constant $c_{\sigma}$ in Eq.~\eqref{eq03c} governs the evolution of
the coupling strengths. The constants
$\sigma_l$ (standing for $\sigma_{\rm low}$), $\sigma_c$ (standing for $\sigma_{\rm center}$),
 and 
$\sigma_h$ (standing for $\sigma_{\rm high}$),  denote the three fixed points of
Eq.~\eqref{eq03c} and, without loss of generality, we assume that $\sigma_l < \sigma_c < \sigma_h$.
In the same equation $s$ is a constant related to the connectivity between 
the coupling of neighboring neurons.  
In this version of link plasticity each neuron/node $i$ has its own time-dependent coupling strength $\sigma_i(t)$
with the neighbors, while the parameters $\mu $, $u_{\rm th}$,  $u_{\rm rest}$,  $R$,
$\sigma_l$, $ \sigma_c $, $ \sigma_h$  and $s$ 
are common to all network elements.

\par As stated in the Introduction,
the adaptivity rule introduced in Eq.~\eqref{eq03c} is related to the evolution of the various link weights
by the influence of their neighboring connectivity. Namely, the evolution of the connectivity $\sigma_i$ is
influenced by the link weights of the neighbors at positions $j$, where $ i-R \le j \le i+R$.
This adaptivity is not to be confused with Hebbian adaptivity rules, where the evolution of link weights
are considered with respect to the 
potential dynamical variables, $u_i$ and $u_j$ \cite{hebb:1949,choi:2022,nunzio:2022,berner:2023}.

\par Biological neurons are known to keep their potential to the rest state for a period of time after firing. 
This inactive period is called {\it refractory period} and is denoted by $T_r$. 
The refractory period is of the order of $T_s$, i.e., it accounts for half the period of the single neurons. 
Earlier studies  have shown that hybrid (chimera or bump) states emerge even 
in the absence of a refractory period \cite{tsigkri:2016,luccioli:2010,olmi:2010,tsigkri:2017}.
To keep the system as simple as possible, and since the refractory period is not essential 
for the development
of chimera or bump states, a refractory period will not be considered in this study. 

\par In the simulations, aiming to keep the system as generic as possible, 
we consider random, homogeneous initial conditions
both for the initial potentials, $u_i(t=0)$ and coupling strengths, $\sigma_i(t=0)$.
 The initial states are drawn randomly
from uniform distributions such that $0 \le u_i(t=0) < u_{\rm th}$ and $-1 \le \sigma_i(t=0)\le 1$.
The system size needs to be chosen large enough, usually $N \ge 1000$ nodes, to approach asymptotic dynamics
avoiding effects related to 
finite system sizes. To treat all elements equally and avoid boundary effects we use periodic boundary 
conditions as also discussed in the Introduction.
The linear chain containing the $N$ oscillators closes forming a ring and all indices are 
considered $\mod N$, $u_i(t)=u_{i+N}(t)$ and similarly $\sigma_i(t)=\sigma_{i+N}(t)$ . Typical spacetime
plots of the network evolution can be viewed in Fig.~\ref{fig:01}a,b and c  for negative fixed points          
$(\sigma_l , \sigma_c , \sigma_h)$ and Fig.~\ref{fig:02}a,b and c  for positive fixed points.         
Detailed description of these patterns and their properties will follow 
in Sec.~\ref{sec:chimera_plasticity} for Fig.~\ref{fig:01} and 
in Sec.~\ref{sec:bumps_plasticity} for Fig.~\ref{fig:02}.

\subsection{Quantitative measures}
\label{sec:quantitative}
\par To quantify the complexity of synchronization patterns induced by the nonlinearity in dynamics together with 
the adaptivity of the bondings
 the most frequently used measure is
the average firing rate, or average frequency, $f_{i}$, 
which counts the number of firings  of neuron $i$ in the unit of time
\cite{omelchenko:2013,omelchenko:2015}.
 Namely, if the oscillator $i$ has performed $Q_{i}$ resettings in the time interval $\Delta T$, the
average firing rate is calculated as:
\begin{equation}
\label{eq04} 
f_{i}= \frac{Q_{i}}{\Delta T}=\frac{1}{T_i}, \>\>\> i=1, \cdots N.
\end{equation}
In Eq.~\eqref{eq04}, $T_i$ is the period of oscillator $i$, which may differ substantially from $T_s$
due to the interactions in the network. The
average firing rates account for the network evolution and dynamics and they can differentiate
between homogeneous and hybrid states only in the case of stable chimeras or bumps. In the case of chimera
states, if the
coherent and incoherent domains 
travel in the network, all network elements spend some amount of time in the coherent 
motion while for other time intervals they perform  incoherent motion. 
This way, for long time averages all elements acquire a common average firing rate. 
The same is true for mobile bumps. This is a main
disadvantage in the use of the firing rates for distinguishing a traveling chimera state from a purely homogeneous 
oscillatory network and similarly for traveling bump states. 

\par In the domain of static and dynamical pattern formation the image entropy has been previously
employed to identify chimera states in complex coupled maps \cite{smidtaite:2019}.   Here, the global entropy, $H(t)$,
is used to identify the evolution and stability of bumps or chimera states. 
$H(t)$ is here defined in terms of the homogeneity/heterogeneity of the coupling strengths as:
\begin{subequations} 
\begin{equation}
H(t)=-\sum_{i}^N\tilde{\sigma}_{i}(t)\log \tilde{\sigma}_{i}(t)
\label{eq05a}
\end{equation}
\begin{equation}
\tilde{\sigma_i}(t)=\frac{\abs{\sigma_i(t) }}{\sum_{j=1}^N \abs{\sigma_i(t)}}.
\label{eq05b}
\end{equation}
\label{eq05}
\end{subequations}
\par In case of negative $\sigma_i$ (inhibitory linking) the absolute values of $\sigma_i$ are used
in Eqs.~\eqref{eq05} to calculate $H(t)$ (and later on in Eq.~\eqref{eq06} for the calculations
of the local entropy).
The global entropy, $H(t)$, can not distinguish 
between moving and immobile chimera and bump states, for the same reasoning as explained
above in the case of the average firing rates. 
\par The local entropy evolution, $H_j(t)$, around node $j$ offers information on how the entropy
changes locally in time and is calculated as:
\begin{equation}
H_j(t)=-\sum_{k=j-R}^{j+R}\tilde{\sigma}_{k}(t)\log \tilde{\sigma}_{k}(t).
\label{eq06}
\end{equation}

\par As in the case of the global entropy, in Eq.~\eqref{eq06} for the calculations
of the local entropy the absolute values of $\sigma_i$ are used if $\sigma_i$ are negative  (inhibitory) linking.

\par Other quantitative measures include the size distribution $P(\sigma )$ of the link weights,
the fraction of elements (relative size of) that belong to 
the active and subthreshold domains \cite{tsigkri:2016,tsigkri:2017}, the distributions 
of firing rates and link weights. As will be discussed in the next sections the quantitative measures
depend on the model parameters, on the connectivity scheme and on the coupling strengths.

\par Without loss of generality, in the next sections the following working parameter set
will be used : 
$\mu =1$, $u_{\rm rest} =0$, $u_{\rm th}=0.98$, $N=1024$ and $R=10$ or $40$. 
The parameters which will be varied for the exploration of the network regimes
are $\sigma_l$, $ \sigma_c $, $ \sigma_h$  and $R$.
For the network integration the forward Euler scheme was used with integration
step is $dt=10^{-3}$ TU. Runge-Kutta integration was also employed in some cases 
for confirmation of the  results.
The integration mostly takes place for 5000 time units (TU).  For the calculation of the firing rates, 
$f_i, \> i=1, \cdots N$, the first 1000 TU are 
considered as transient and are ignored.

\section{The formation of composite chimera-like states due to multistability in the plasticity}
\label{sec:chimera_plasticity}

\par
\begin{figure}[t]
\includegraphics[width=0.32\textwidth,angle=0.0]{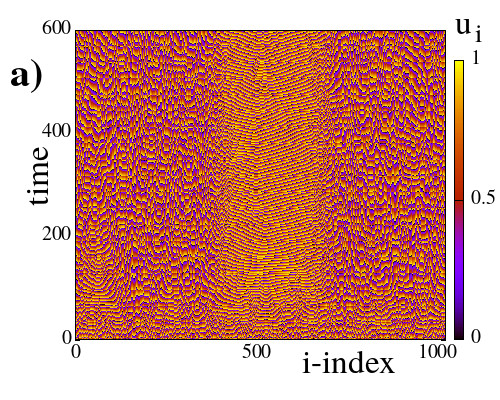}
\includegraphics[width=0.32\textwidth,angle=0.0]{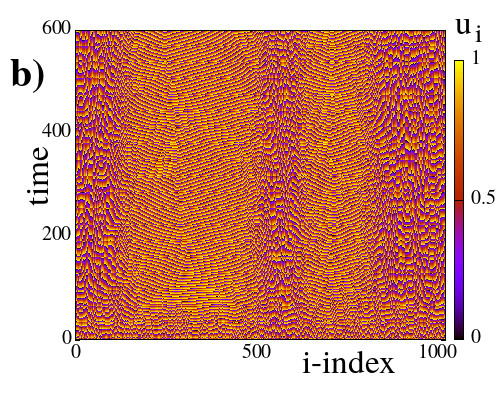}
\includegraphics[width=0.32\textwidth,angle=0.0]{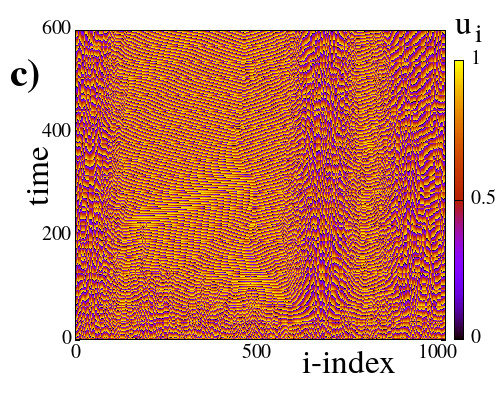}\\
\hspace*{0.1cm}
\includegraphics[width=0.32\textwidth,angle=0.0]{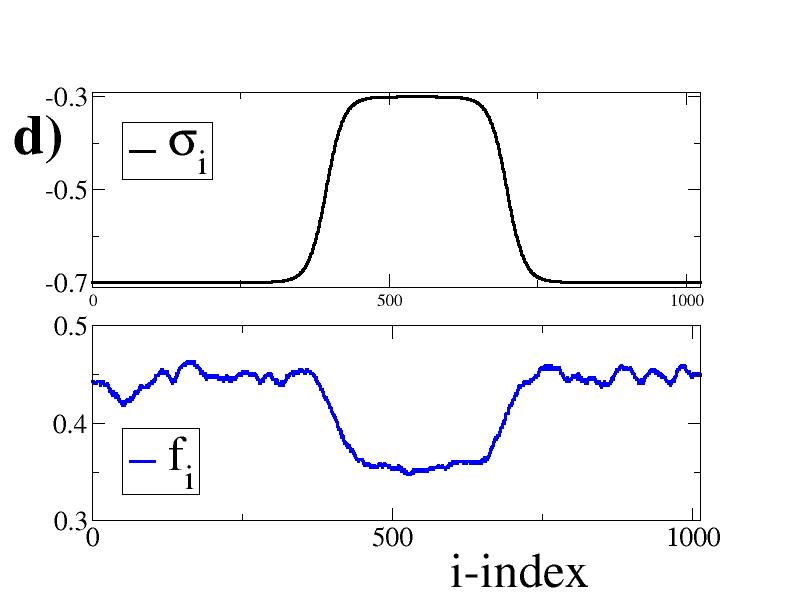}
\includegraphics[width=0.32\textwidth,angle=0.0]{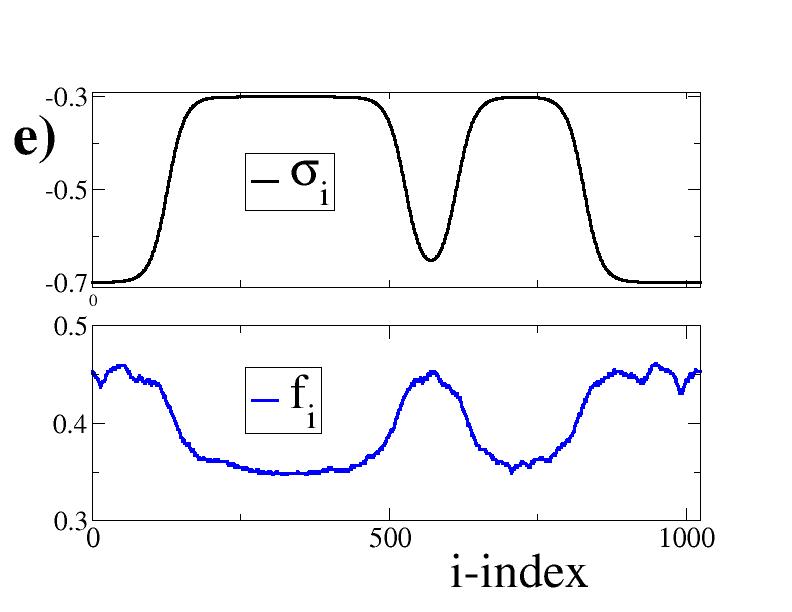}
\includegraphics[width=0.32\textwidth,angle=0.0]{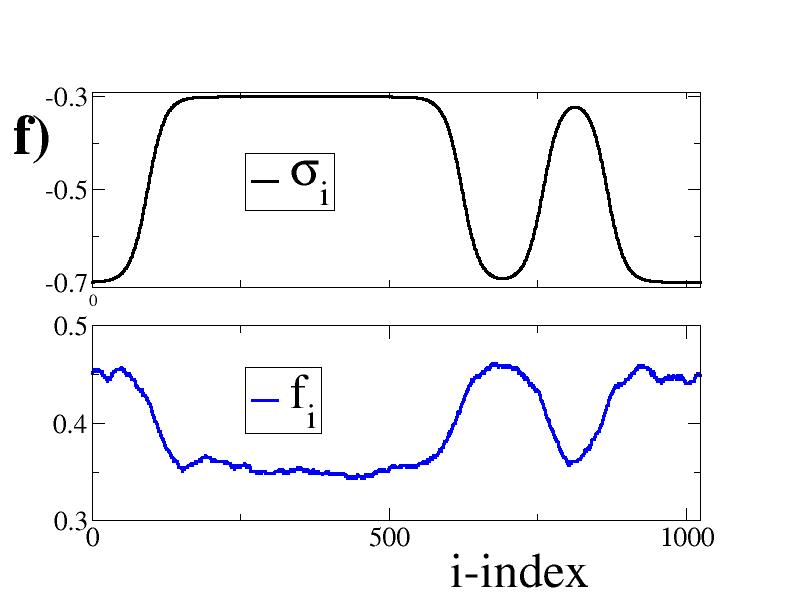}\\
\includegraphics[width=0.50\textwidth,angle=0.0]{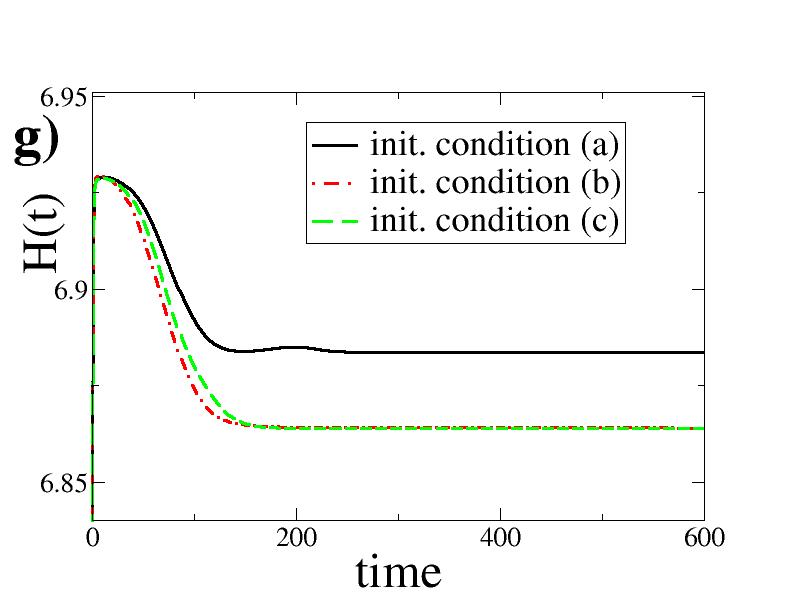}
\includegraphics[width=0.50\textwidth,angle=0.0]{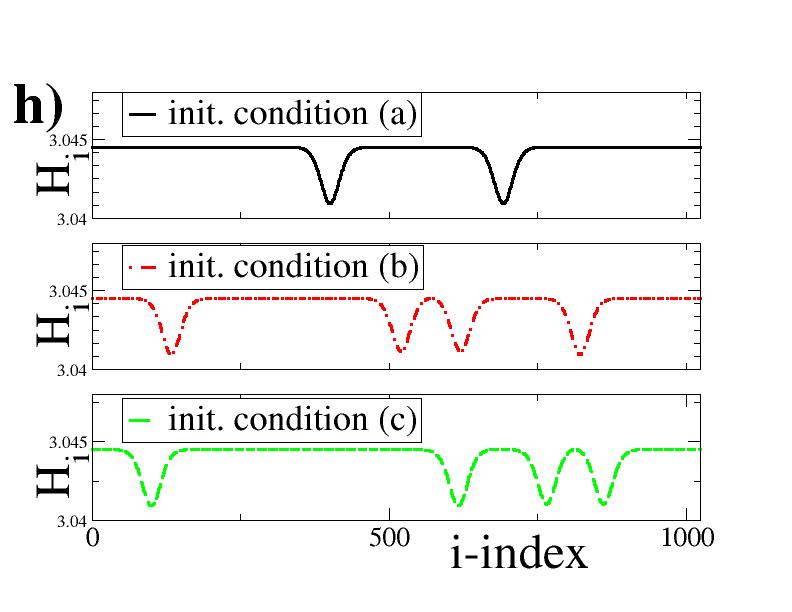}
    \caption{\label{fig:01}  Coupled LIF dynamics with bistable plasticity
and negative coupling strengths. Top row: spacetime plots starting from three
different initial conditions are depicted in panels a), b) and c).
Middle row : d), e) and f) are the corresponding asymptotic coupling strengths and firing rates
for the above three initial conditions. Bottom row:
g) The entropy evolution with time, $H(t)$ and h) The local entropy values at the final stages of
the simulations. In panels g) and h)
the black solid lines correspond to initial condition a), the
red dashed-dotted line to b) and the green dashed line to c).
The simulations in a), b) and c) start from different random initial 
conditions in $u_i$ and $\sigma_i$.
All other parameter values are identical in the three cases: 
$\mu =1$, $u_{\rm rest} =0$, 
$u_{\rm th}=0.98$, $N=1024$, $R=10$, $\sigma_l =-0.7$, $ \sigma_c =-0.5$,
 $ \sigma_h=-0.3$, $c_\sigma =-1.0$ and $s=0.9$.
}
\end{figure}

 As first examples we plot in Figs.~\ref{fig:01}a, \ref{fig:01}b and \ref{fig:01}c the spacetime
plot of the system, Eqs.~\eqref{eq03}, starting with three different initial
conditions both in the $u_i$ and $\sigma_i$ variables. All other system and
network parameters are the same for all nodes. After the transient states, the three initial
conditions drive the system to distinct final states which contain domains
with different firing rates and $\sigma$-distributions. In particular, in 
Fig.~\ref{fig:01}a the system develops one large domain of low firing rates
coexisting with a smaller domain of high firing rates. In Figs.~\ref{fig:01}b
and \ref{fig:01}c the systems develop two domains of low firing rates coexisting with
two domains of high firing rates, the size of the different domains being
variable in panels \ref{fig:01}b and \ref{fig:01}c. The above simulations indicate that the 
introduction of plasticity in the links introduces a kind of memory effects
because the final hybrid states develop domains of different sizes,
reflecting the particularities of the initial distribution of the $\sigma_i$
values. We need to stress here that the fixed points $(\sigma_l , \sigma_c , \sigma_h)=(-0.7, -0.5, -0.3)$ 
used in the present simulations take all negative values, giving rise to 
oscillatory dynamics. Because the different $\sigma$-values influence  
locally the system dynamics we observe chimera-like states. 
Note that in the absence of link-weight evolution (no plasticity,
$c_{\sigma}=s=0$), the final
chimera states of the LIF model develop coherent and incoherent domains whose 
sizes do not depend on the initial conditions and only the position of 
the domains may be displaced in the network \cite{tsigkri:2016,tsigkri:2018}.
For a quick comparison of the LIF network evolution with and without coupling 
plasticity, we add in the Appendix \ref{sec:appendixA} typical spacetime
plots of the system evolution under different coupling parameters as
indicated on the plots.

\par Regarding the firing rates,
for  the three different initial conditions shown in Figs.~\ref{fig:01}a, \ref{fig:01}b and \ref{fig:01}c 
the firing rates $f_i$ and the coupling strength $\sigma_i$ distributions are shown in 
panels ~\ref{fig:01}d, \ref{fig:01}e and \ref{fig:01}f, respectively. The spatial distributions of  $f_i$ and $\sigma_i$
also reflect the formation of domains in the system, corroborating the observations in the
spacetime plots related to the memory effects, as discussed in Sec.~\ref{sec:adaptive}.

\par These memory effects are further reflected in the evolution of the
global entropy which also depends strongly
on the initial conditions. As an example, we plot in Fig.~\ref{fig:01}g 
the entropy evolution in the three
cases of identical parameters while starting from different initial conditions. 
We observe that each initial condition (a, b or c) leads to a final state 
with different entropy. Note that in the case of constant and equal $\sigma_i$ values, then
$\tilde{\sigma}_i=1/N$. Consequently, a constant value of the entropy is expected
only in the case of equipartition and in this case 
$H$ only depends on the system size as $H=\log N$. For the present system,
$N=1024$ and therefore $H=6.93$ for constant and equal $\sigma_i$, independently
of the system parameters. We note that during transient all initial conditions
reach $H$-values close to 6.93 but at the final state they all drop considerably
below 6.9, indicating certain degrees of organization.

\par In panel h of Fig.~\ref{fig:01} we plot the local entropy $H_i$ at the
final state of the system for the three initial conditions. We may observe that
the local entropy drops at the points that correspond to transitions between the
different domains. Using Eq.~\eqref{eq06} for the case of constant values of
$\sigma_i$ in different local domains of the system, when the domain size is
$2R+1$ the local entropy has the value $H_i=\log (2R+1)$. For $R=10$ used in
Fig.~\ref{fig:01}, the local entropy values become  $H_i=3.0445$. Indeed,
such values appear in regions of constant $\sigma_i$ values, see the
corresponding areas in panels  \ref{fig:01}a, \ref{fig:01}b and \ref{fig:01}c 
as well as in \ref{fig:01}d, \ref{fig:01}e and \ref{fig:01}f.

\par In regions of transition between domains of different $\sigma$-values the local
entropy drops, as demonstrated by the variations in the $H_i$ in panel \ref{fig:01}h.
These drops delineate the borders between the different domains of coherence,
while the number of entropy maxima or minima can be used to quantitatively
count the number of coherent/incoherent domains in the chimera state.
The local entropy can be also viewed as an indicator of plasticity and is also linked
to the memory of the initial distribution of $\sigma$-values.
\par The results in the present section and in Appendix \ref{sec:appendixA}
are all obtained under negative (inhibitory) coupling strengths. 
As will be discussed in the next section,
Sec.~\ref{sec:bumps_plasticity}, and in Appendix \ref{sec:appendixB}, 
in the cases of positive coupling strengths (eg., $(\sigma_l , \sigma_c , \sigma_h)=(0.3, 0.5, 0.7)$ 
fixed points) bump and bump-like states will be observed.
\section{The formation of composite bump-like states due to multistability in the plasticity}
\label{sec:bumps_plasticity}

\par
\begin{figure}[h]
\includegraphics[width=0.32\textwidth,angle=0.0]{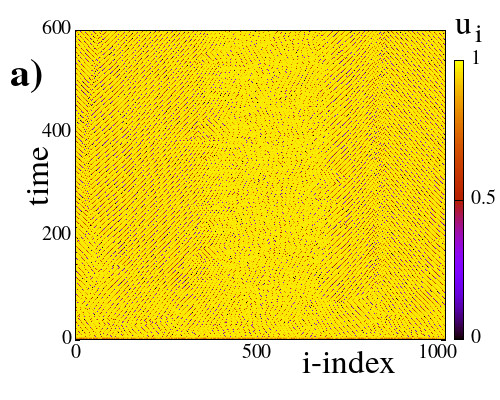}
\includegraphics[width=0.32\textwidth,angle=0.0]{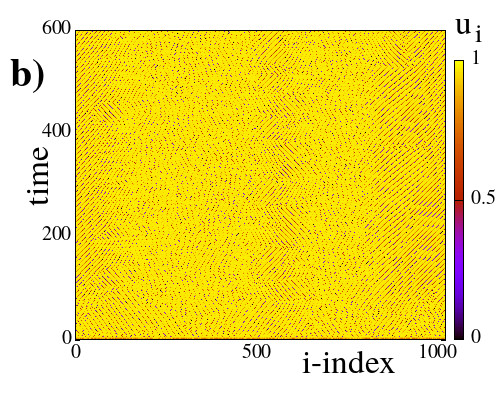}
\includegraphics[width=0.32\textwidth,angle=0.0]{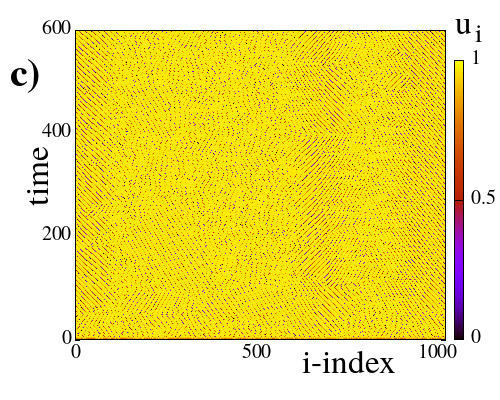}\\
\includegraphics[width=0.32\textwidth,angle=0.0]{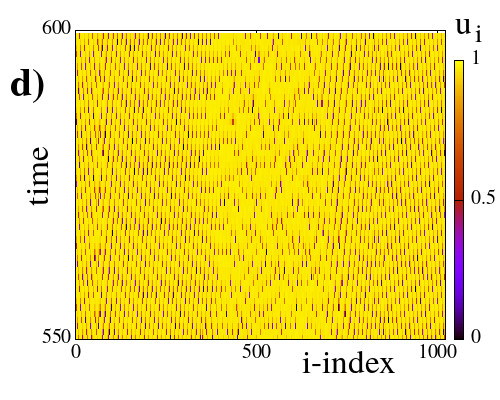}
\includegraphics[width=0.32\textwidth,angle=0.0]{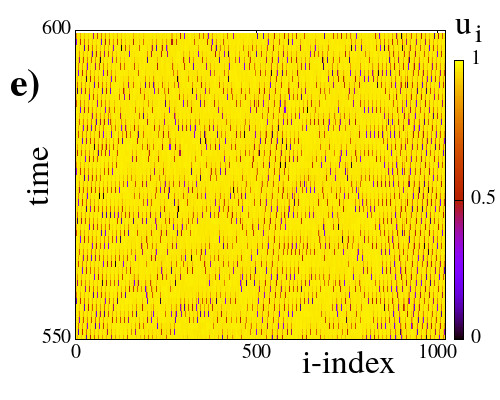}
\includegraphics[width=0.32\textwidth,angle=0.0]{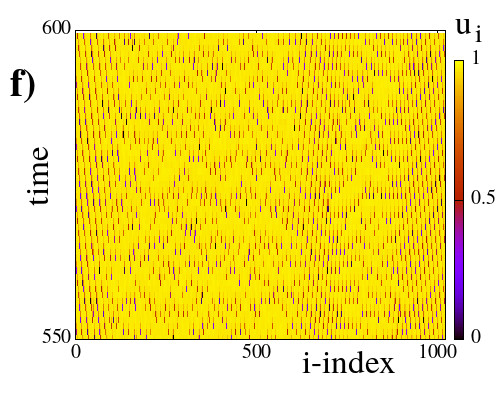}\\
\hspace*{0.1cm}
\includegraphics[width=0.32\textwidth,angle=0.0]{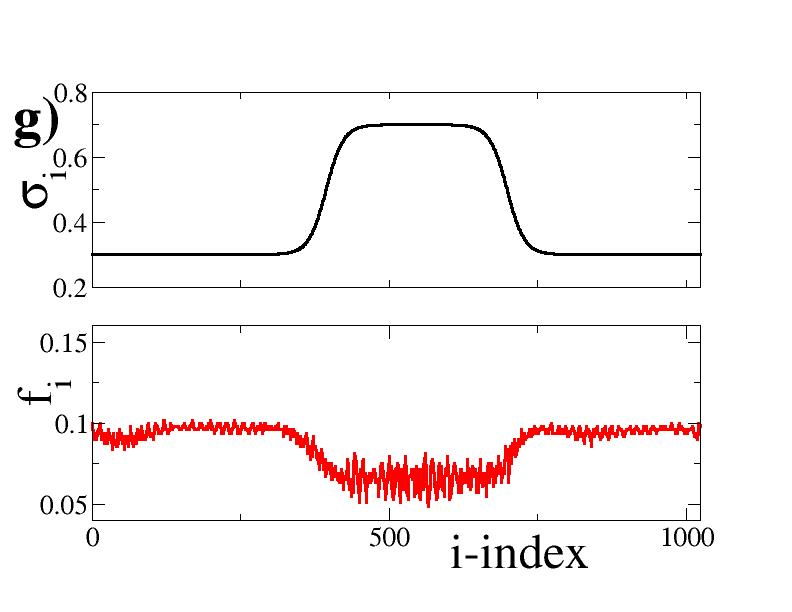}
\includegraphics[width=0.32\textwidth,angle=0.0]{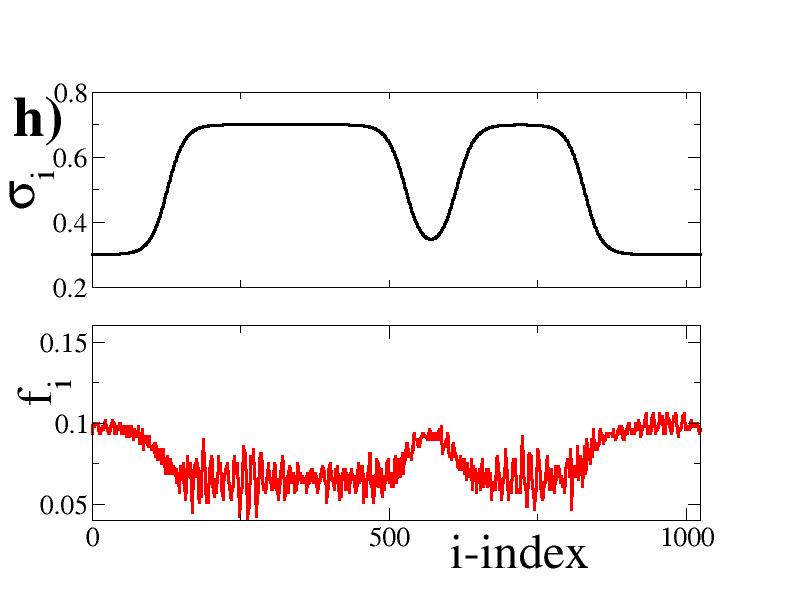}
\includegraphics[width=0.32\textwidth,angle=0.0]{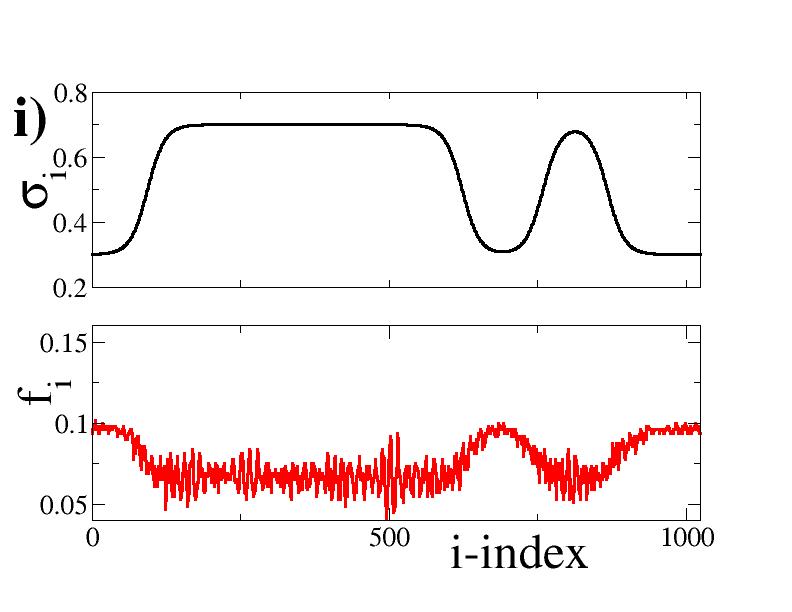}\\
\includegraphics[width=0.50\textwidth,angle=0.0]{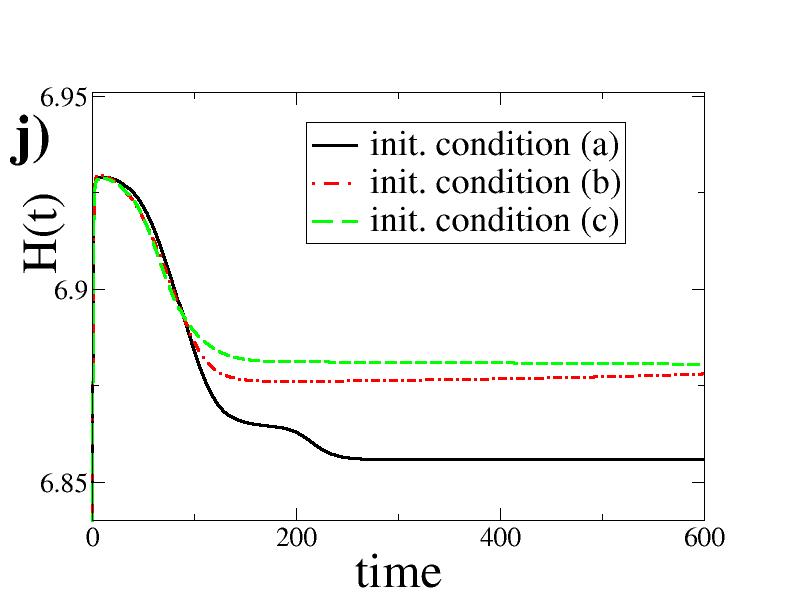}
\includegraphics[width=0.50\textwidth,angle=0.0]{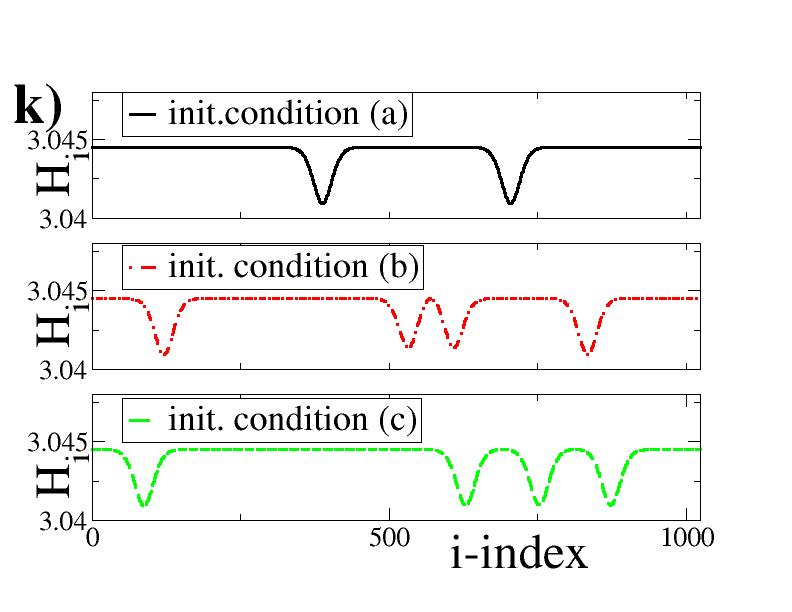}
    \caption{\label{fig:02} Coupled LIF dynamics with bistable plasticity
and positive coupling strengths. Typical potential spacetime profiles $u_i$ for three
different initial conditions a), b) and c)
forming bump-like states. For better understanding of the
the bump state complexity in the second row
panels d), e) and f) show specific
details of panels a), b) and c), respectively.
Third row : g), h) and i) are the corresponding coupling strengths and firing rates
for the above three initial conditions.
Forth row: j) The entropy evolution in time, $H(t)$ and k) The local entropy values at the final stages of
of the simulations. In panels j) and k)
the black solid lines correspond to initial condition a), the
red dashed-dotted line to b) and the green dashed line to c).
The simulations in a), b) and c) start from different random initial 
conditions in $u_i$ and $\sigma_i$.
Coupling fixed point values are:  $\sigma_l =+0.3$, $\sigma_c =+0.5$ and $\sigma_h =+0.7$.
All other parameter values are the same as in Fig.~\ref{fig:01}.
}
\end{figure}

 In this section we report on the effects
of multistable plasticity in the case of positive fixed points $\sigma_h$,
$\sigma_c$ and $\sigma_l$ in the dynamics of the coupling strengths.
In parallel to the case of negative (inhibitory) couplings, here
 we perform numerical integration of Eqs.~\eqref{eq03} for 
positive (excitatory) coupling strengths with
$\sigma_l=+0.3$, $\sigma_c=+0.5$ and $\sigma_h=+0.7$, starting with three different
 random initial 
conditions in $u_i$ and $\sigma_i$. All other parameters are identical to the ones in 
Fig.~\ref{fig:01} following the working parameter set. The results are shown
in Fig.~\ref{fig:02}.

\par The top row of Fig.~\ref{fig:02} presents the spacetime plot of the
potentials $u_i$ for three different random initial 
conditions in $u_i$ and $\sigma_i$ with all other
parameters identical. The time interval is here extended from $t=0$ to 
600 TUs and covers
the transient time. In all three cases, Figs.~\ref{fig:02}a, \ref{fig:02}b and \ref{fig:02}c, the network nodes spend
considerable time near the threshold value (yellow regions) while only occasionally
they oscillate and reset to the rest state. This activity is characteristic
of bump states, see also Ref.~\cite{tsigkri:2016} and Appendix~\ref{sec:appendixB}.
Because the bump states here (Fig.~\ref{fig:02}) are formed under adaptivity conditions,
where asymptotically the system has reached a state of variable coupling weights, these states
will be called ``bump-like'' states. Moreover, these bump-like states
consist of domains where different average
firing rates dominate on a silent background of subthreshold
elements (yellow regions in Figs.~\ref{fig:02}a - \ref{fig:02}f ).

\par For better inspection of the
life and death of the occasional activity in the system, in Figs.~\ref{fig:02}d,
\ref{fig:02}e and \ref{fig:02}f we present details of the panels \ref{fig:02}a, \ref{fig:02}b and \ref{fig:02}c, respectively.
In the 2nd row the time interval is restricted to the last 50 time units,
$t=550-600$TU. The different activity regimes are hard to discern in panels \ref{fig:02}a
- \ref{fig:02}f but they are better visible in the firing rate and coupling strength
diagrams. Figure panels~\ref{fig:02}g, \ref{fig:02}h and \ref{fig:02}i present the coupling strength  and the
firing rate spatial distributions starting with the different random initial conditions.
 We note that in the case of Fig.~\ref{fig:02} where coupling strengths 
are positive the firing rates are
considerably lower than in Fig.~\ref{fig:01} where the coupling strengths 
are negative. These lower firing rates  are attributed to
the tendency of the nodes to stay at subthreshold values for long time
intervals and to occasionally perform resettings to the rest state.
This has also previously been reported  in the case of non-adaptive couplings
and for positive coupling
strengths, see Ref.~\cite{tsigkri:2017} and Appendix \ref{sec:appendixB}.

\par Memory effects are also evident here. Comparing Figs.~\ref{fig:02}a,
\ref{fig:02}b and \ref{fig:02}c and corresponding quantitative indices  Figs.~\ref{fig:02}g,
\ref{fig:02}h and \ref{fig:02}i, we understand that the initial (random) spatial distributions
of $\sigma_i$ have given rise to steady states with different domain
sizes, influenced by the specific initial $\sigma_i(t=0)$ states. We recall that in case of
identical linking sizes ($\sigma_i=\sigma = const.$ and without adaptivity) and different initial
conditions $u_i(t=0)$, when chimera states or bump states are formed
 the distribution of domain sizes remain statistically constant,
independent of the randomly chosen initial conditions, $u_i(t=0)$. 

\par Regarding the entropy evolution in the network, in panel~\ref{fig:02}j we present
the evolution of the global entropy in time for the three different initial
conditions. We observe similar approaches to the steady state as in the
case of inhibitory dynamics but here the entropy levels attained are
different and the first initial condition shows
the lowest final entropy (compare with Fig.~\ref{fig:01}g ).

\par With respect to local entropy, changes in the $H_i$ values occur
at the transition regions between active bumps and subthreshold domains,
see panel~\ref{fig:02}k.
Similarly to the case of inhibitory coupling, the drops in the local
entropy permit to count (quantitative index) the number of active bumps
and subthreshold domains of the bump pattern. As in Sec.~\ref{sec:chimera_plasticity},
the local entropy profile also retains some characteristics of the original
spatial distribution of the links (memory effects). 
In addition, adaptivity has
caused the shift of bump state formation toward shorter coupling ranges
$R$ (in Fig.~\ref{fig:02} $R=10$). This shift in the formation of complex
synchronization patterns along with the shifts shown in the case of
chimera-like states will be further discussed in Sec.~\ref{sec:coupling-range}.

\section{Coexistence of chimeras and bumps due to multistability in the plasticity}
\label{sec:multistable_plasticity}

\par
\begin{figure}[h]
\includegraphics[width=0.52\textwidth,angle=0.0]{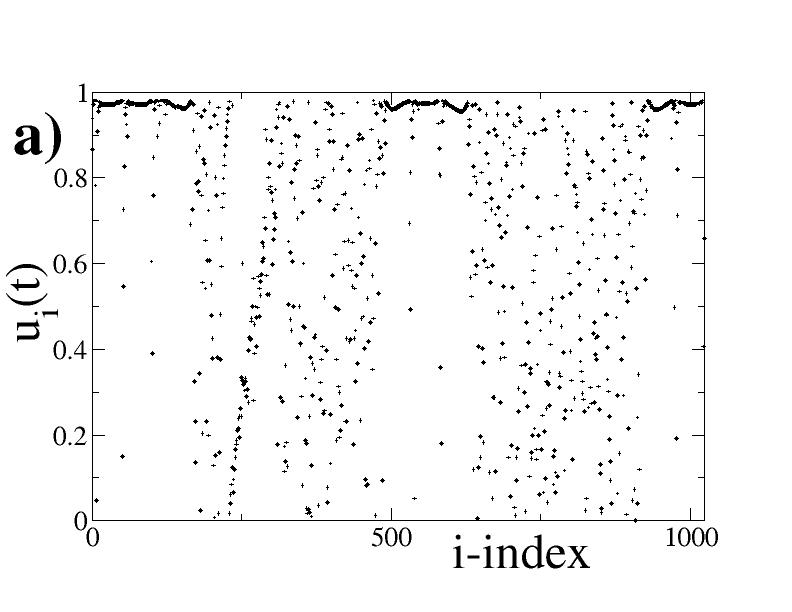}
\hspace{-0.4 cm}
\includegraphics[width=0.53\textwidth,height=0.37\textwidth,angle=0.0]{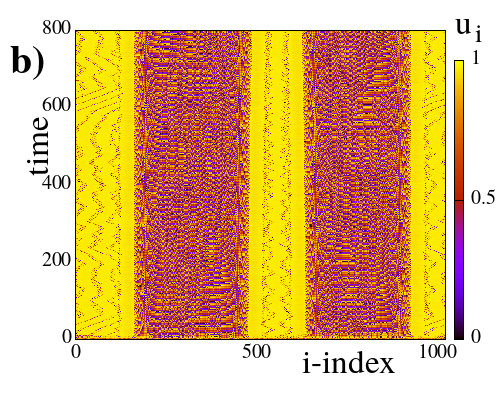}\\
\includegraphics[width=0.52\textwidth,angle=0.0]{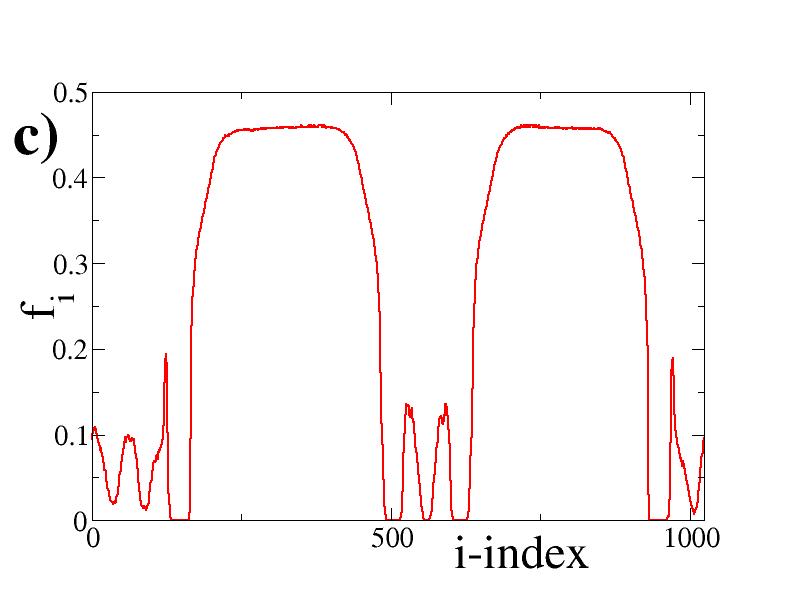}
\includegraphics[width=0.52\textwidth,angle=0.0]{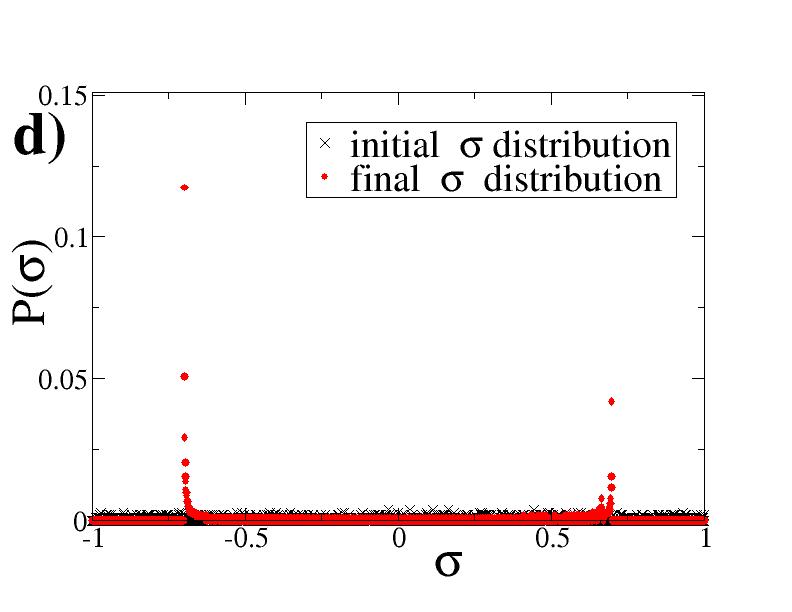}
    \caption{\label{fig:03} LIF network with bistable adaptive coupling
comprising positive and negative fixed points cause the formation of complex synchronization patterns combining 
both bump and chimera features. a) Typical snapshot of the neuron potential 
(time 800 TU), b) spacetime plot c) the average firing rate after 800 TU
and d) The distribution of coupling strengths $\sigma$ at $t=0$ TU (black
crosses, homogeneous random initial conditions) and at $t=800$ TU (red
dots). Parameter values are: $\mu =1$, $u_{\rm rest} =0$, 
$u_{\rm th}=0.98$, $N=1024$, $R=40$, $\sigma_l =-0.7$, $ \sigma_c =0.0$,
 $ \sigma_h=+0.7$, $c_\sigma =-1.0$ and $s=0.9$.
 Simulations start from random uniform initial 
conditions in $u_i$ and $\sigma_i$.
}
\end{figure}

 Starting with random coupling strengths distributed on the various nodes
of the network (in addition to random initial potentials $u_i(t=0)$), 
we now concentrate on an exemplary case where the fixed points
in the dynamics of the coupling weights (Eq.~\eqref{eq03c})
take mixed positive and negative negative values. Mixed excitatory
and inhibitory links are well known to coexist in the same network, notably in the brain dynamical
networks \cite{haddad:2014}.
Without loss of generality, we use the working parameter set (see subsection \ref{sec:quantitative}),
with appropriate values of $\sigma_l$, $ \sigma_c $, $ \sigma_h$  and $s$. 
The plots in Fig.~\ref{fig:03} are representative complex hybrid patterns in the
presence of synaptic plasticity with mixed positive and negative fixed points. 


\par More specifically, in Fig.~\ref{fig:03} we present typical results of the complex
hybrid patterns produced when synaptic plasticity is considered in the coupling strength
with $\sigma_l =-0.7$, $ \sigma_c =0.0$,
and $ \sigma_h=+0.7$. Previous studies have revealed that for negative (inhibitory) coupling strengths
chimera states are supported by the LIF network \cite{tsigkri:2016,tsigkri:2018}, while bumps states appear for
positive (excitatory) couplings \cite{tsigkri:2017}. By imposing bifurcation conditions in the
coupling during the
evolution of the LIF network we record hybrid behavior as indicated in Fig.~\ref{fig:03}a. In this
panel we note coexistence of two incoherent domains with domains where subthreshold oscillations persist.
Within the subthreshold domains we can record small active bumps. This is further clarified in the
spacetime plot of Fig.~\ref{fig:03}b; here we may first note the presence of yellow subthreshold regions
supporting small bumps which move stochastically to the left and right on the ring. The subthreshold regions
are separated by active incoherent domains characterized by different $u_i$ values. The positions of
the active incoherent and the subthreshold domains on the ring stay fixed in time. 
In fact, the erratically traveling
small bumps in the yellow (subthreshold) regions
do not cross the incoherent domains, but they are scattered back by them. In Fig.~\ref{fig:03}c
we record the firing rates and we note that the active domains demonstrate the highest firing frequencies,
while the bumps show lower frequencies.  The difference in the firing frequencies between bumps and the incoherent
domains may be attributed both to the occasional firing and to the motion of the former ones.
As the bumps move the energy is transmitted 
from one element to the neighbors and the firing changes position in time. This means that the elements
spend part of their time at the subthreshold state and part in the active state. Because the firing rates
are recorded as averages over many time units, they are underestimated in the case of the small mobile bumps.
As a result we record lower firing activity inside the subthreshold regions on the positions of the bumps.
In Fig.~\ref{fig:03}d, we plot with black crosses the initial distribution $P(\sigma )(t=0)$ of the coupling strengths
$\sigma_i$ and with red dots the long time (asymptotic) steady state distribution, $P(\sigma )(t=800)$. 
Because the coupling strengths were
initially chosen randomly and homogeneously in the interval (-1.0, 1.0) the initial distribution 
of $\sigma_i$ is flat.
At the steady state of the network the distribution $P(\sigma )$ becomes bimodal, with maxima at  $\sigma \sim \sigma_l=-0.7$
and $\sigma \sim \sigma_h=0.7$, as formatted by the bistable Eq.~\ref{eq03c}.  
\par Another observation worth mentioning is the difference in height
in the two maxima of the final bimodal distribution $P(s)$. Indeed, the maximum amplitude corresponding to 
inhibitory coupling, $P(\sigma \sim -0.7)=0.12$, is higher than the excitatory coupling one, $P(\sigma \sim +0.7)=0.04$.
This is not unexpected, since the active chimera-like regions cover a larger number of nodes than the 
active bumps in the subthreshold regions. Which of the two domains (chimera-like regions or subthreshold regions)
 will dominate in the final state is, again, a matter
of initial conditions.

\par To the best of our knowledge, complex synchronization patterns composed by bump-like and chimera-like states
co-habitating on the network is a direct effect of adaptivity and have not been observed before.


\section{The influence of the coupling range in synaptic bistability}
\label{sec:coupling-range}
\par As earlier discussed, one of the effects of synaptic bistability
is the presence of chimera or bump states for small values of the coupling
range $R$. We also recall that in these parameter regions ($R <<$) spatially hybrid
states are not formed for constant or homogeneous linking, 
see also Appendices~\ref{sec:appendixA} and~\ref{sec:appendixB}. 
To explore the $R$-regions where chimera-like or bump-like states are formed
due to synaptic bistability
we perform numerical simulations of LIF networks for inhibitory and
excitatory coupling and $R$ ranging from 1 to 80 lattice units.
For quantifying the presence of hybrid states we use the 
deviation of the local entropy at the asymptotic state, $d^2_H$, defined as:
\begin{subequations} 
\begin{equation}
d^2_H=\frac{1}{N}\sum_{j=1}^{N}\left[H_{\rm max}-H_j\right] ^2.
\label{eq07a}
\end{equation}
\begin{equation}
H_{\rm max}= \max \{ H_j \}, \>\>\> j=1\cdots N.
\label{eq07b}
\end{equation}
\label{eq07}
\end{subequations} 
\noindent In Eqs.~\eqref{eq07} the values of $H_i$ and $H_{\rm max}$ are
recorded at the asymptotic state, after the transient period.
\par
\begin{figure}[h]
\centering
\includegraphics[width=0.8\textwidth,angle=0.0]{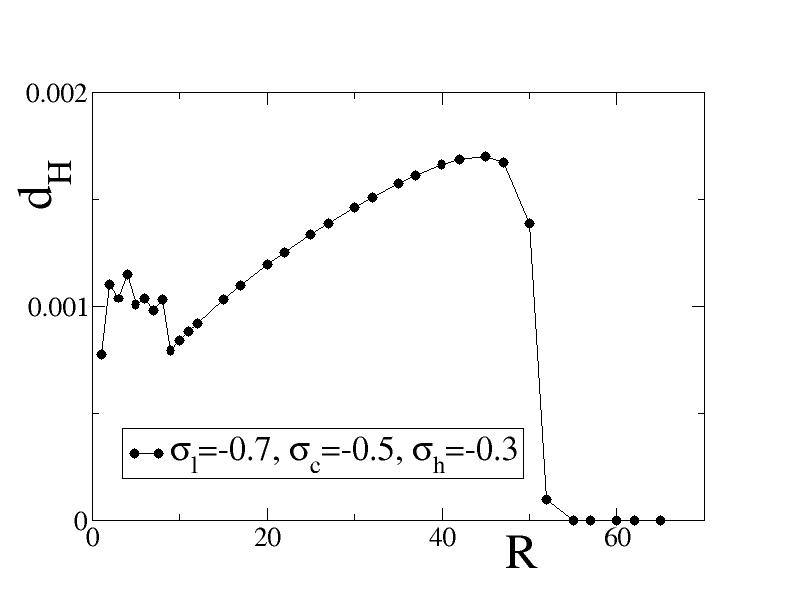}
    \caption{\label{fig:04} Local entropy deviations $d_H$ as a function
of the coupling range $R$ for inhibitory coupling. 
 Parameter values are: $\sigma_l =-0.7$, $ \sigma_c =-0.5$,
 $ \sigma_h=-0.3$. Other parameters are as in Fig.~\ref{fig:01}.
 All simulations start from the same random uniform initial 
conditions both in $u_i$ and in $\sigma_i$ .
}
\end{figure}

 In Fig.~\ref{fig:04} we present variations of $d_H$ with $R$ 
for inhibitory coupling, using as coupling fixed points
the exemplary  set  $(\sigma_l , \sigma_c , \sigma_h )$ = $(-0.7,-0.5,-0.3)$.
The fixed points in the selected set take all negative values
leading to inhibitory asymptotic dynamics.
Other parameters as in the working parameter set.
The solid black line with black circles in Fig.~\ref{fig:04} serves as guide for the eyes.

\par For small values of the coupling range, $R \le 10$, we observe non-monotonous
variations of $d_H$ as a function of $R$. We note that for $R = 1$ we have the limit
of diffusion where hybrid synchronization regimes are
not observable. In this short range limit, $1 \le R \le 10$, 
the coupling ranges are very short and any small increase in
$R$ may influence substantially the spatial distribution
of domains where $\sigma_l$ or $\sigma_h$ dominate. That is the reason of 
observing the non-monotonous variations in $d_H$ .
\par For intermediate coupling distances we observe a monotonous 
increase of $d_H$ with $R$.  As $R$ increases organization takes place in larger and larger domains and
this is reflected in the index $d_H$. For even larger values
of $R$, one of the two stable fixed points prevails and the spatial
distribution of link weights becomes homogeneous.
As a result all weights become equal and, therefor, $d_H \to 0$.
The shape of  $d_H$ vs. $R$ curve in Fig.\ref{fig:04} is typical for general
inhibitory parameter values but the
$R$-value where the transition from bistability to monostability 
occurs depends on the precise values of the
fixed points $(\sigma_l , \sigma_c , \sigma_h )$. For the 
exemplary set (-0.7,-0.5,-0.3)
the transition occurs at $R \sim 55$, while for (-0.8,-0.5,-0.1)
at $R \sim 90$ (not shown). These transition values may also vary slightly
for different initial conditions. The difference in the $R$
value where the transition occurs can be dependent also on
the length of the interval between the attracting fixed
points, which for the case of set (-0.7,-0.5,-0.3) is 0.4,
while for set (-0.8,-0.5,-0.1) is 0.7.

\par
\begin{figure}[h]
\centering
\includegraphics[width=0.8\textwidth,angle=0.0]{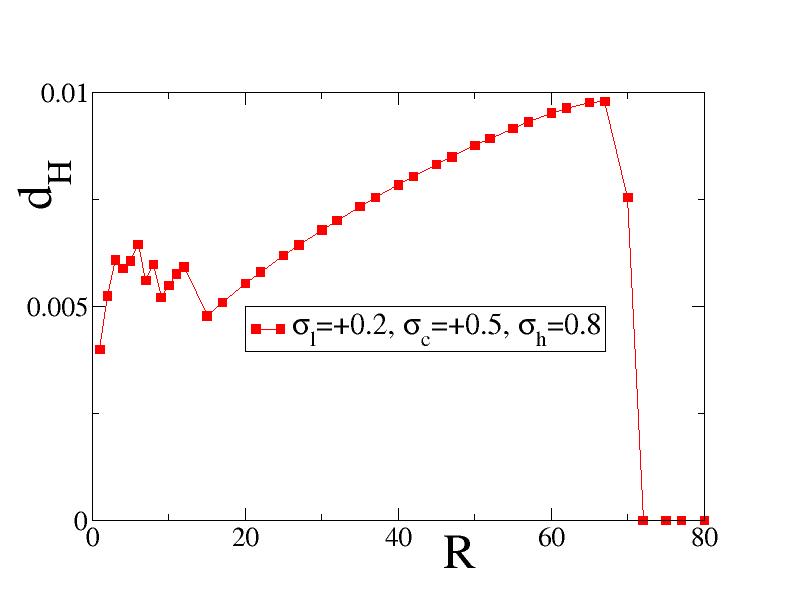}
    \caption{\label{fig:05} Local entropy deviations as a function
of the coupling range for excitatory coupling. 
 Parameter values are: $\sigma_l =+0.2$, $ \sigma_c =+0.5$,
 $ \sigma_h=+0.8$ (red squares-line). Other parameters are as in Fig.~\ref{fig:01}.
 All simulations start from the same random uniform initial 
conditions both in $u_i$ and in $\sigma_i$.
}
\end{figure}

 In Fig.~\ref{fig:05} we examine the case of positive fixed points
and present variations of the local entropy deviation $d_H$ with
the coupling range $R$
for excitatory coupling. As  exemplary set of coupling fixed points we use
$(\sigma_l , \sigma_c , \sigma_h )$=(+0.2, 0.5,+0.8)
(the solid red line is set as guide for the eyes). 
Other parameters as in Fig.~\ref{fig:04}.
\par As in the case of Fig.~\ref{fig:04}, in Fig.~\ref{fig:05} we also  observe non-monotonous variations
of $d_H$ vs. $R$ in the small coupling ranges, $1\le R\le 17$.
The reason is similar to the case of negative fixed points, namely,
for small $R$-values any increase in $R$ may
influence substantially the spatial distribution of domains where $\sigma_l$ or
$\sigma_h$ dominate. 
Similarly, for intermediate coupling ranges we observe a monotonic increase
of $d_H$ vs. $R$ up to an $R$-value where transition occurs and
one of the two fixed points, $\sigma_l$ or $\sigma_h$, dominates, 
leading again to an homogeneous
link-weight distribution. For the present parameter values
the critical value of $R$ where the transition occurs is 74. 
Apart from the difference in the transition coupling range, another
difference between Figs.~\ref{fig:04} and~\ref{fig:05} is
in the scales of $d_H$ values: the $d_H$ in Fig.~\ref{fig:05} are
approximately one order of magnitude higher than those in Fig.~\ref{fig:04}.
\par Comparing Fig.~\ref{fig:04} and Fig.~\ref{fig:05},
we conclude that the general form of the $d_H$ vs. $R$ curve is 
essentially conserved for positive or negative adaptive couplings.
While the form of the curve is conserved 
for an extended region of parameter values, 
the precise $R$ and $d_H$ ranges where the non-monotonous behavior
or monotonous increase dominate as well as the $R$-values 
where $d_H$ drops to 0 (transition point) depend strongly on the specific 
coupling parameter values and in particular in the values and
the signs of the three fixed points  $(\sigma_l,  \sigma_c, \sigma_h)$.

\section{Conclusions}
\label{sec:conclusions}

\par In the present study we concentrate on the influence of bistable plasticity rules
in the dynamics of LIF networks. Previous studies have demonstrated
the presence of chimera states for inhibitory coupling strengths
and bump states for excitatory couplings in nonlocally connected LIF networks. 
In the case of inhibitory coupling, we show that multistable
synaptic rules may shift (lower) considerably the region of coupling ranges 
where chimera states are observed. Similarly, for the
positive couplings, the introduction of bistability in the
synaptic  strengths  lowers considerably the $R$ ranges where bumps occur.
In addition, we show that for appropriate choice of the stable
synaptic fixed points it is possible to obtain cohabitation of
chimera-like and bump-like states simultaneously on the network.

\par To quantify the presence of chimera/bump states in the network we use
 the local entropy  deviations $d_H$ in the network. Using
$d_H$ as an index of synaptic organization, we show that there is 
a transition point in the coupling range values $R$ where the system
transits from multistable to single fixed point dynamics.
While the form of  $d_H$ vs $R$ retains its qualitative features for
large regions of the parameters, 
the transition $R$-values depend strongly on the specific coupling parameter values.

\par Memory effects have also been recorded, where the spatial organization
of the final state
of the network contains recollection of the original $(t=0)$ distributions 
of link weights. 
Therefore, starting from different initial conditions we may end up in 
final states with statistically different spatial ordering/organization,
eventhough all system parameters are kept to the same values.

\par Direct extensions of this work may include the study of the
$R$-transition as a function of the distance between the stable fixed points
$\sigma_l$ and $\sigma_h $ or the coupling strength $s$,
 as well as various aspects of cohabitation
of chimera and bump states simultaneously on the network.

\par Future extensions may include the possibility of multistable linking
 $\sigma_{ij}(t)$ which depends both on the pre- and post-synaptic neurons.
Other,  Hebbian-like plasticity rules \cite{hebb:1949,choi:2022} 
or spike-timing-dependent plasticity \cite{caporale:2008}
may include nonlinear linking terms depending
on the values of the involved potentials $u_i$ and $u_j$
 on the last term of the right-hand side in Eq.~\eqref{eq03c}.
In a different direction, yet, we may consider 
the influence of a power law
distribution on the number of links emanating from the neurons of the
network or/and power-law distribution of the link weights. 
This conforms with recent biomedical studies reporting that the
number of synapses per axon follow long-tailed distributions \cite{lin:2024}
and is in line with Hebb's propositions of long-range distributions
in the links per node \cite{hebb:1949}. 
This structural property has been
recently attributed to the property of preferential attachment of the neurons in the network
\cite{scheffer:2020}. 

\section{Acknowledgments}
This work was 
supported by computational time granted from the Greek Research \& Technology Network (GRNET)
in the National HPC facility - ARIS - under Project ID: PR014004.

\begin{appendices}
\renewcommand{\theequation}{A.\arabic{equation}}
\setcounter{equation}{0}
\setcounter{figure}{0}                       
\renewcommand\thefigure{A.\arabic{figure}}   
\par In the two appendices below we recapitulate some results from earlier publications
on the LIF model without adaptation rules. Basically, we integrate Eqs.~\eqref{eq03a}
and Eq.~\eqref{eq03b}, while keeping all coupling strengths $\sigma_i$ to a constant value $\sigma$.
The constant $\sigma$'s are
negative in the case of inhibitory coupling (see Appendix Sec.~\ref{sec:appendixA}) and
positive in the case of excitatory coupling (see Appendix Sec.~\ref{sec:appendixB}).
These results may be found scattered in previous publications, notably in \cite{tsigkri:2016,tsigkri:2018,tsigkri:2017}.
Here, we provide a short recapitulation of these results
for the convenience of the readers and for direct comparison
of the LIF networks with and without adaptation rules. 

\par The equations which
now describe the system without adaptation are:
\begin{subequations}
\begin{equation}
\label{eq-appA1a} 
\frac{du_{i}(t)}{dt}= \mu - u_{i}(t)+\frac{\sigma}{2R}\sum_{j=i-R }^{i+R} 
\left[ u_{j}(t) - u_{i}(t)\right]
\end{equation}
\begin{equation}
\lim_{\epsilon \to 0}u_{i}(t+\epsilon ) \to u_{\rm rest}, 
\>\>\> {\rm when} \>\> u_{i}(t) \ge u_{\rm th}.
\label{eq-appA1b}
\end{equation}
\label{eq-appA1}
\end{subequations}
With respect to Eq.~\eqref{eq03a}, Eqs.~\eqref{eq-appA1a} has coupling
strength $\sigma$ which is constant in time and takes a value common
to all nodes. Eq.~\eqref{eq03c} is redundant (not needed) here, since all coupling
strengths  neither evolve nor adapt in time.

\section{Results for LIF network without adaptation rule: Inhibitory Coupling}
\label{sec:appendixA}
\par In this section we present the system evolution without coupling plasticity and for negative
values of the coupling strength (inhibitory coupling). For comparative reasons
we will present spacetime plots for $\sigma=-0.3$ and $-0.7$ with $R=10$ and $R=350$.
All other parameters are the same as the ones used in the main text.
\par
\begin{figure}[h]
\includegraphics[width=0.5\textwidth,angle=0.0]{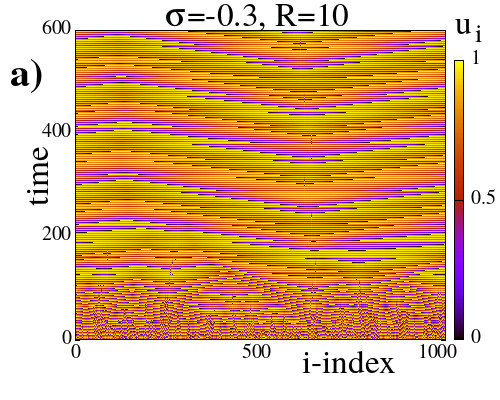}
\includegraphics[width=0.5\textwidth,angle=0.0]{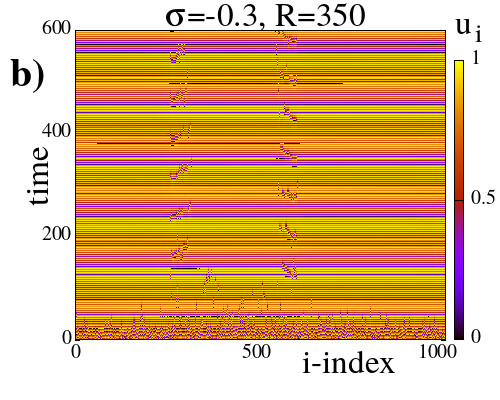}
\includegraphics[width=0.5\textwidth,angle=0.0]{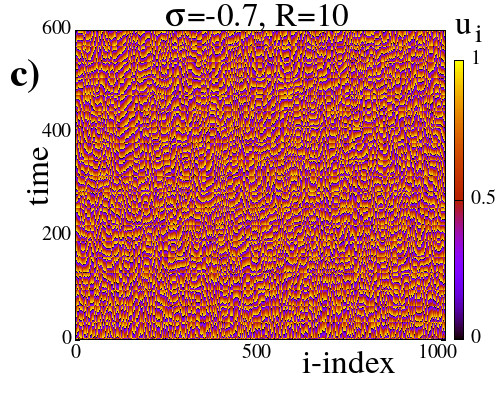}
\includegraphics[width=0.5\textwidth,angle=0.0]{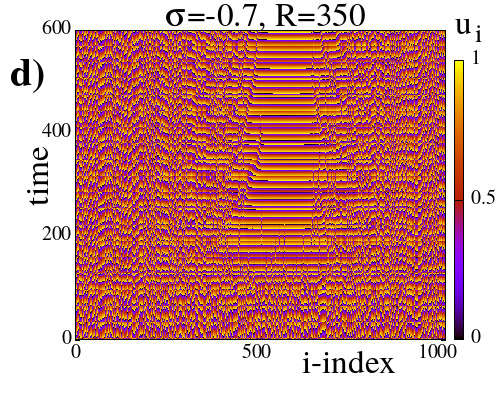}
    \caption{\label{fig-Append1} Synchronization patterns for inhibitory
coupling without link plasticity.  Typical spacetime plots
for a) $\sigma =-0.3$ and $R=10$, b) $\sigma =-0.3$ and $R=350$, c) 
$\sigma =-0.7$ and $R=10$, and d) $\sigma =-0.7$ and $R=350$. 
All other parameter values are
as in Fig.~\ref{fig:01} of the main text.
 Simulations start from the same random initial conditions in $u_i(t=0)$.
}
\end{figure}

 In Fig.~\ref{fig-Append1}a we present the spacetime plot of the system
for $\sigma=-0.3$  corresponding to the first of the fixed points
in the coupling strength (as in Sec.~\ref{sec:chimera_plasticity} and Fig.~\ref{fig:01}) and $R=10$. 
We see that for homogeneous and short-distance coupling 
the system, after transient, reaches the homogeneous steady state. For these
couplings strengths ($\sigma=-0.3$) only for long-distance couplings non-trivial
effects arise such as coexistence of coherent and incoherent domains,
see  Fig.~\ref{fig-Append1}b where $\sigma=-0.3$ and $R=350$.
[Nevertheless, careful observations of Fig.~\ref{fig-Append1}a may spot the presence of a single solitary state
around $i=650$, where the node escapes the  altogether coherent environment. Solitary states
are related to the birth of chimera states.]
\par Similarly, in Fig.~\ref{fig-Append1}c we present the spacetime plot of the system
for smaller value $\sigma=-0.7$ corresponding to the second attractive fixed point
in the coupling strength presented in Sec.~\ref{sec:chimera_plasticity} and Fig.~\ref{fig:01}. 
The coupling range is small, $R=10$, as in Fig.~\ref{fig-Append1}a
and in the other cases of plasticity studied in the main text. We note a certain degree
of inhomogeneous firing in the system without indications of coherence. For this
coupling strength typical chimera states develop when the coupling range
takes large enough values,  see  Fig.~\ref{fig-Append1}d with $\sigma=-0.7$ and $R=350$.
For a more complete study on the presence of chimera states for negative coupling strengths
and different parameter values in the LIF model without plasticity 
we refer the interested reader to Refs.~\cite{tsigkri:2016,tsigkri:2018}.

\section{Results for LIF network without adaptation rule: Excitatory Coupling}
\label{sec:appendixB}

\setcounter{figure}{0}                       
\renewcommand\thefigure{B.\arabic{figure}}   
\par In analogy with the presentation in the previous appendix section~\ref{sec:appendixA}, we here recall results
on the system evolution without coupling plasticity and for positive
values of the coupling strength (excitatory coupling).
\par
\begin{figure}[h]
\includegraphics[width=0.5\textwidth,angle=0.0]{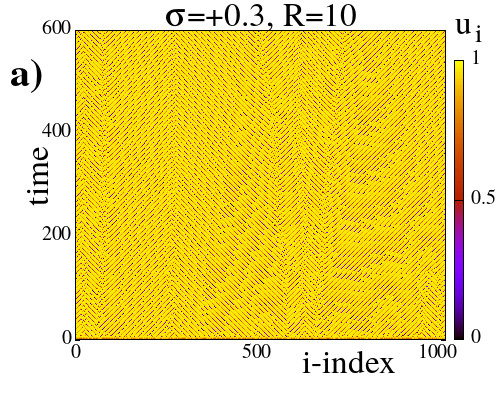}
\includegraphics[width=0.5\textwidth,angle=0.0]{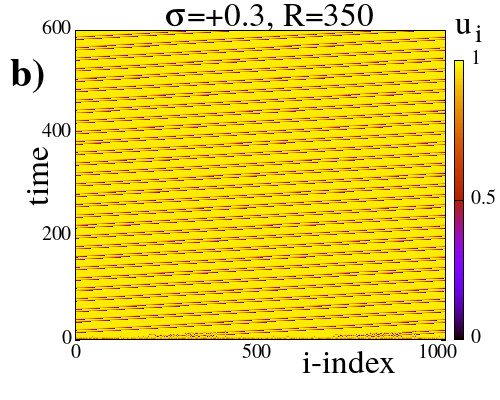}
\includegraphics[width=0.5\textwidth,angle=0.0]{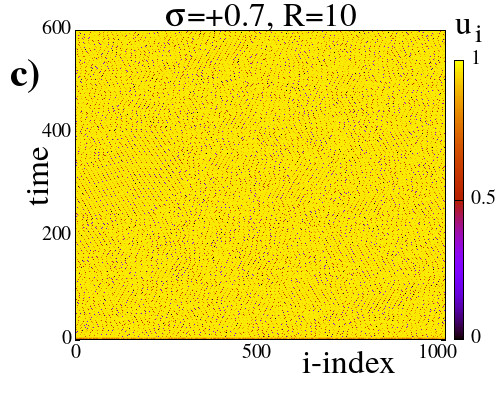}
\includegraphics[width=0.5\textwidth,angle=0.0]{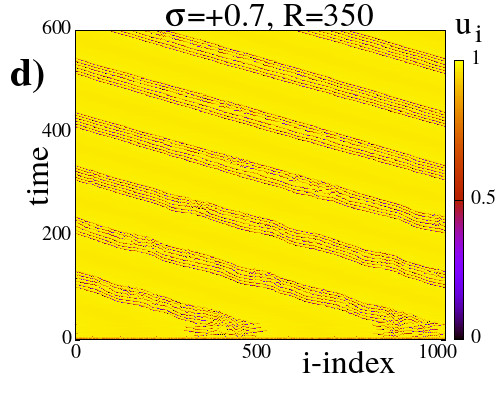}
    \caption{\label{fig-Append2} Synchronization patterns for inhibitory
coupling without link plasticity.  Typical spacetime plots
for a) $\sigma =0.3$ and $R=10$, b) $\sigma =0.3$ and $R=350$, c) 
$\sigma =0.7$ and $R=10$, and d) $\sigma =0.7$ and $R=350$. 
All other parameter values are
as in Fig.~\ref{fig:01} and in appendix \ref{sec:appendixA}, Fig.~\ref{fig-Append1}.
 Simulations start from the same random initial conditions in $u_i(t=0)$.
}
\end{figure}

 In this case ($\sigma >0$ identical for all nodes), the system presents bump states where 
active elements appear in a background of silent subthreshold elements.
Namely, for small values of the coupling range we observe erratically
appearing and moving active elements, see the cases of $R=10$ in
Figs.~\ref{fig-Append2}a and~\ref{fig-Append2}c for $\sigma=+0.3$ and $+0.7$,
respectively. As $R$ increases, the active nodes organize in domains
which travel around the ring with constant velocity. The size of the
active traveling domains and their velocity depend on the parameters
$R$ and $\sigma$, while their direction (left-wise or right-wise) is
determined by the initial conditions. Typical examples of traveling bumps
are presented in Figs.~\ref{fig-Append2}b and~\ref{fig-Append2}d.
These figures indicate that the size of the active traveling bumps
increases with $\sigma$. Further discussions on studies on bump states
in 1D LIF networks without adaptation are presented in Ref. \cite{tsigkri:2017}.

\end{appendices}

\bibliographystyle{unsrt}
\bibliography{./plasticity2024.bib}

\end{document}